\documentclass[10pt,twocolumn]{article}

\usepackage[utf8]{inputenc}
\usepackage[T1]{fontenc}
\usepackage{lmodern}
\usepackage{microtype}
\usepackage{eurosym}
\usepackage{url}

\usepackage{fontawesome5}
\usepackage{colortbl}
\usepackage{multirow}
\usepackage{lipsum}
\usepackage{amsthm}
\usepackage{mathtools}
\usepackage{xfrac}
\usepackage[export]{adjustbox}
\usepackage{longtable}
\usepackage{makecell}
\usepackage{booktabs}
\usepackage[table,xcdraw]{xcolor}
\usepackage{amssymb}
\usepackage{graphicx}
\usepackage{subcaption}
\usepackage{verbatim}
\usepackage{placeins}
\usepackage{mdframed}
\usepackage{hyperref}
\usepackage{fancyvrb}
\usepackage{bera}
\usepackage{pdfpages}
\usepackage{pgf-umlsd}
\usepackage{tikz}
\usepackage[square,numbers]{natbib}
\usepackage{fancyhdr}
\usepackage{lastpage}
\usepackage{dirtree}
\usepackage{forest}
\usepackage{wrapfig}
\usepackage{algorithm}
\usepackage{algorithmic}
\usepackage{listings}
\usepackage{geometry}

\definecolor{cai_primary}{HTML}{4C9A99}  
\definecolor{cai_secondary}{HTML}{307FE2}  
\definecolor{cai_accent}{HTML}{1D8348}  
\definecolor{cai_dark}{HTML}{3F4444}  
\definecolor{cai_light}{HTML}{F5F5F5}  
\definecolor{cai_purple}{HTML}{8A4FFF}  

\definecolor{unitree_primary}{HTML}{4C9A99}
\definecolor{unitree_red}{HTML}{E74C3C}
\definecolor{unitree_orange}{HTML}{E67E22}
\definecolor{unitree_green}{HTML}{27AE60}

\hypersetup{
    colorlinks=true,
    urlcolor=cai_primary,
    linkcolor=cai_primary,
    filecolor=cai_primary,
    citecolor=cai_primary,
    pdftitle={Cybersecurity AI: Humanoids as Attack Vectors},
    pdfauthor={Alias Robotics Research}
}
\renewcommand{\headrulewidth}{0.4pt}
\renewcommand{\footrulewidth}{0.4pt}
\renewcommand{\headrule}{\hbox to\headwidth{\color{cai_dark!30}\leaders\hrule height \headrulewidth\hfill}}
\renewcommand{\footrule}{\hbox to\headwidth{\color{cai_dark!30}\leaders\hrule height \footrulewidth\hfill}}

\usepackage{caption,setspace}
\usepackage{siunitx}
\usepackage{pdflscape}
\usepackage{enumitem}
\setlist{itemsep=0pt,topsep=0pt,parsep=0pt,partopsep=0pt,leftmargin=*}

\usepackage{titlesec}
\titleformat{\section}
  {\normalfont\normalsize\bfseries\color{cai_primary}}
  {\thesection}
  {0.5em}
  {}
  [\titlerule]

\titleformat{\subsection}
  {\normalfont\footnotesize\bfseries\color{cai_accent}}
  {\thesubsection}
  {0.5em}
  {}

\titleformat{\subsubsection}
  {\normalfont\footnotesize\bfseries\color{cai_dark}}
  {\thesubsubsection}
  {0.5em}
  {}

\titlespacing*{\section}{0pt}{3pt}{2pt}
\titlespacing*{\subsection}{0pt}{2pt}{1pt}
\titlespacing*{\subsubsection}{0pt}{1pt}{1pt}

\usepackage[normalem]{ulem}

\usepackage{authblk}
\usepackage{fancyvrb}  

\renewcommand\Affilfont{\small\normalfont}
\definecolor{cai_affil_color}{HTML}{3F8984}

\makeatletter
\renewcommand\AB@affilsepx{\\\protect\Affilfont}
\let\orig@maketitle\maketitle
\renewcommand{\maketitle}{%
  \orig@maketitle%
  \vspace{-1.5em}%
  {\color{cai_primary!30}\hrule height 0.5pt}%
  \vspace{1em}%
}
\makeatother

\title{\LARGE\textcolor{cai_primary}{\textbf{Cybersecurity AI: Humanoid Robots as Attack Vectors}}}
\date{} 

\author[1]{Víctor Mayoral-Vilches}
\author[3]{Andreas Makris}
\author[2]{Kevin Finisterre}

\affil[1]{
    {\normalfont\textcolor{cai_primary}{\textbf{Alias Robotics}}\\
    {\tt\footnotesize\textcolor{cai_primary}{\faEnvelope}~victor@aliasrobotics.com \quad \textcolor{cai_primary}{\faGlobeEurope~\href{https://aliasrobotics.com}{aliasrobotics.com}} }}    
}

\affil[2]{
    {\normalfont\textcolor{cai_primary}{\textbf{Confidential}}
}}

\affil[3]{
    {\normalfont\textcolor{cai_primary}{\textbf{thinkAwesome! GmbH}}
}}

\affil[*]{
    {\normalfont{\faGithub}~{\tt\footnotesize \href{https://github.com/aliasrobotics/cai}{https://github.com/aliasrobotics/cai}}}\\
    {\normalfont{\faDiscord}~{\tt\footnotesize \href{https://discord.gg/fnUFcTaQAC}{https://discord.gg/fnUFcTaQAC}}}
}

\begin{document}

\frenchspacing

\twocolumn[
\begin{@twocolumnfalse}
\maketitle
\vspace{-1em}
\begin{abstract}
\footnotesize

We present a systematic security assessment of the Unitree G1 humanoid showing it operates simultaneously as a covert surveillance node and can be purposed as an active cyber operations platform. Initial access can be achieved by exploiting the BLE provisioning protocol which contains a critical command injection vulnerability allowing root access via malformed Wi-Fi credentials, exploitable using hardcoded AES keys shared across all units. Partial reverse engineering of Unitree's proprietary FMX encryption reveal a static Blowfish-ECB layer and a predictable LCG mask—enabled inspection of the system's otherwise sophisticated security architecture, the most mature we have observed in commercial robotics. Two empirical case studies expose the critical risk of this humanoid robot: (a) the robot functions as a trojan horse, continuously exfiltrating multi-modal sensor and service-state telemetry to 43.175.228.18:17883 and 43.175.229.18:17883 every 300 seconds without operator notice, creating violations of GDPR Articles 6 and 13; (b) a resident Cybersecurity AI (CAI) agent can pivot from reconnaissance to offensive preparation against any target, such as the manufacturer's cloud control plane, demonstrating escalation from passive monitoring to active counter-operations. These findings argue for adaptive CAI-powered defenses as humanoids move into critical infrastructure, contributing the empirical evidence needed to shape future security standards for physical–cyber convergence systems.

\end{abstract}
\vspace{0.2cm}

\vspace{0.1cm}
\end{@twocolumnfalse}
]

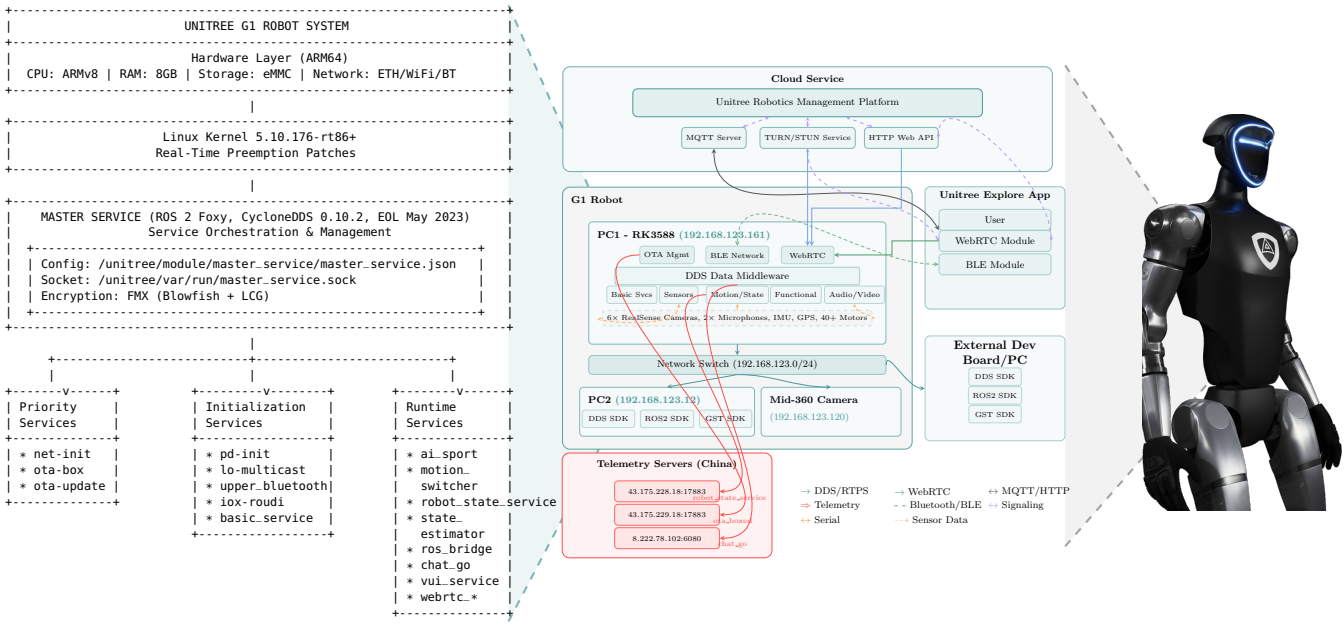
\begin{figure*}[h!]
  \centering
  \begin{tikzpicture}[overlay, remember picture]
      \coordinate (left-start) at (-2.2cm, 0cm);
      \coordinate (right-end) at (0cm, -5cm);
      \coordinate (left-start-2) at (-2.2cm, -8.2cm);
      \coordinate (right-end-2) at (0cm, -4cm);

      \fill[unitree_primary, opacity=0.1]
          (left-start) -- (right-end) -- (right-end-2) -- (left-start-2) -- cycle;

      \draw[dashed, thick, unitree_primary, opacity=0.7]
          (left-start) -- (right-end);
      \draw[dashed, thick, unitree_primary, opacity=0.7]
          (left-start-2) -- (right-end-2);

      \coordinate (right-start) at (5.2cm, -0.8cm);
      \coordinate (left-end) at (8cm, -5cm);
      \coordinate (right-start-2) at (5.2cm, -7.2cm);
      \coordinate (left-end-2) at (8cm, -4cm);

      \fill[gray, opacity=0.1]
          (right-start) -- (left-end) -- (left-end-2) -- (right-start-2) -- cycle;

      \draw[dashed, thick, gray, opacity=0.7]
          (right-start) -- (left-end);
      \draw[dashed, thick, gray, opacity=0.7]
          (right-start-2) -- (left-end-2);
  \end{tikzpicture}

  \begin{subfigure}[c]{0.2\textwidth}
      \centering
      \begin{Verbatim}[fontsize=\tiny]
+---------------------------------------------------------------------+
|                        UNITREE G1 ROBOT SYSTEM                      |
+---------------------------------------------------------------------+
|                         Hardware Layer (ARM64)                      |
|  CPU: ARMv8 | RAM: 8GB | Storage: eMMC | Network: ETH/WiFi/BT       |
+---------------------------------------------------------------------+
                                  |
+---------------------------------------------------------------------+
|                     Linux Kernel 5.10.176-rt86+                     |
|                    Real-Time Preemption Patches                     |
+---------------------------------------------------------------------+
                                  |
+---------------------------------------------------------------------+
|    MASTER SERVICE (ROS 2 Foxy, CycloneDDS 0.10.2, EOL May 2023)     |
|                   Service Orchestration & Management                |
|  +--------------------------------------------------------------+   |
|  | Config: /unitree/module/master_service/master_service.json   |   |
|  | Socket: /unitree/var/run/master_service.sock                 |   |
|  | Encryption: FMX (Blowfish + LCG)                             |   |
|  +--------------------------------------------------------------+   |
+---------------------------------------------------------------------+
                                  |
      +---------------------------+---------------------------+
      |                           |                           |
+-------v------+          +---------v--------+        +--------v------+
| Priority     |          | Initialization   |        | Runtime       |
| Services     |          | Services         |        | Services      |
+--------------+          +------------------+        +---------------+
| * net-init   |          | * pd-init        |        | * ai_sport    |
| * ota-box    |          | * lo-multicast   |        | * motion_     |
| * ota-update |          | * upper_bluetooth|        |   switcher    |
+--------------+          | * iox-roudi      |        | * robot_state_service
                          | * basic_service  |        | * state_      |
                          +------------------+        |   estimator   |
                                                      | * ros_bridge  |
                                                      | * chat_go     |
                                                      | * vui_service |
                                                      | * webrtc_*    |
                                                      +---------------+
      \end{Verbatim}
  \end{subfigure}
  \hfill
  \begin{subfigure}[c]{0.45\textwidth}
      \centering
      \includegraphics[width=\textwidth]{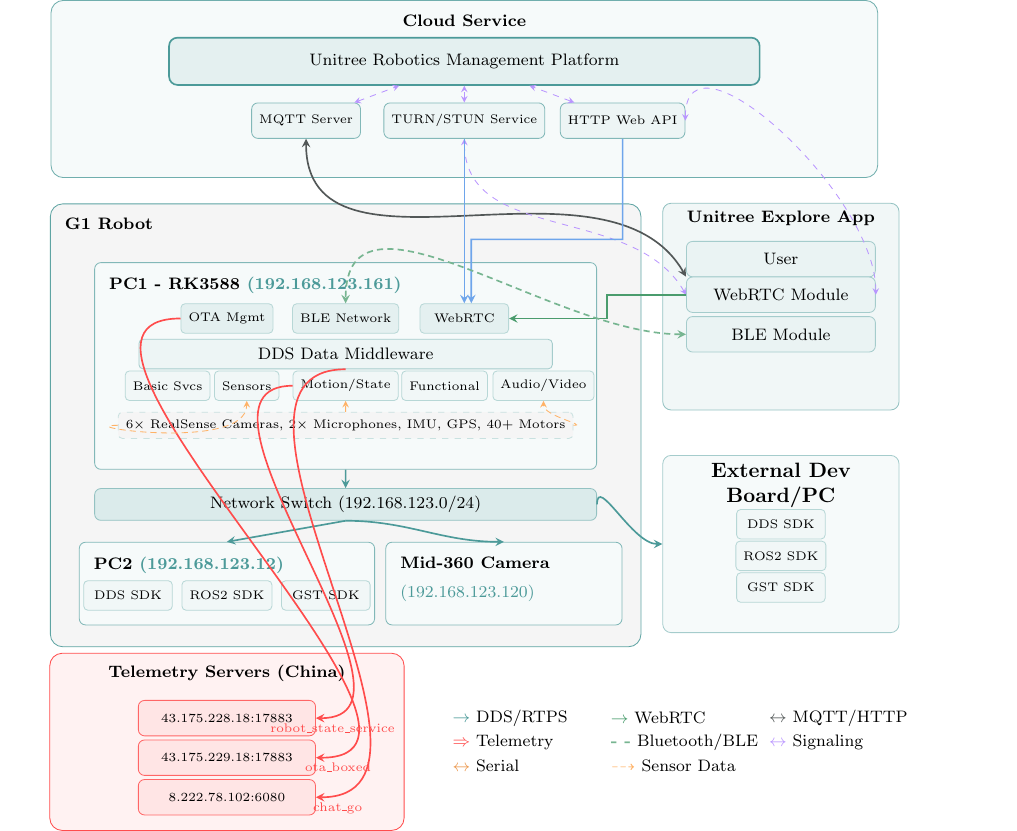}
  \end{subfigure}
  \begin{subfigure}[c]{0.15\textwidth}
    \centering
    \includegraphics[width=\textwidth]{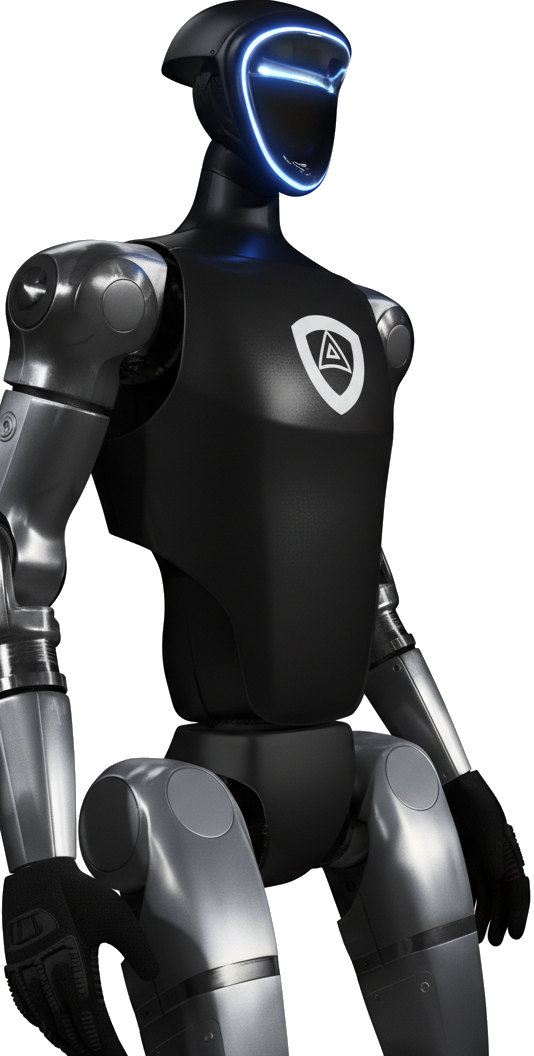}
\end{subfigure}

  \caption{Humanoid Robot Architecture: Internal system structure showing hardware layer, Linux kernel, master service orchestration, and service hierarchy (left), and high-level ecosystem with communication paths showing authorized cloud services, telemetry servers, and internal components including obstacle avoidance, path planning, and speech recognition with DDS/ROS2 compatibility (right). \textcolor{red!70}{Critical finding: Persistent telemetry connections to external servers transmit robot state and sensor data without explicit user consent.}}
  \label{fig:hl_arch}
\end{figure*}

\footnotesize

\section{Introduction}

This technical brief summarizes the findings from Alias Robotics' security assessment in combination with researchers Andreas Makris$^3$ and Kevin Finisterre$^2$ independent findings\footnote{\label{fn:unipwn}Refer to \url{https://github.com/Bin4ry/UniPwn} for Andreas Makris' and Kevin Finisterre's exploitation of this vulnerability.} of the Unitree~G1 humanoid robot. Refer to the full report \cite{mayoralvilches2025cybersecurityhumanoidrobot} and for more details. Our methodology builds upon established prior robotics security assessment practices~\cite{mayoralrvd, mayoral2020devsecops, mayoral2021hacking, mayoral2022alurity,maggi2022security, mayoral2022robot, mayoral2022robothackingmanual, mayoral2025offensive, quarta2017experimental} combining static firmware analysis, binary reverse engineering of the 9.2~MB \texttt{master\_service} orchestrator, runtime network traffic analysis, and BLE protocol exploitation to systematically evaluate the platform's security architecture and more contemporary research on Cybersecurity AI \cite{aliasrobotics2025cai, mayoralvilches2025cybersecurityaidangerousgap, mayoralvilches2025caifluencyframework, mayoral2025cybersecurity}. Five critical findings emerged:
\begin{itemize}
    \item \textbf{Disovered the FMX encryption, which exhibits fundamental cryptographic weaknesses.} The dual-layer scheme employs Blowfish-ECB with a static 128-bit key (effective entropy: 0 bits due to fleet-wide key reuse across all devices) combined with a partially reverse-engineered LCG obfuscation layer (limited to 32-bit seed space). This violates Kerckhoffs's principle—security relies on key secrecy, not algorithm obscurity~\cite{schneier1996applied}.
    \item \textbf{BLE provisioning enables trivial remote compromise.} The Wi-Fi configuration protocol accepts unvalidated input through BLE channels, allowing command injection via SSID/password fields using the pattern \textcolor{unitree_primary}{\texttt{";\$(cmd);\#"}}. Combined with hardcoded AES-CFB keys (\textcolor{unitree_primary}{\texttt{df98b715d5c6ed2b25817b6f2554124a}}) shared across all G1, H1, and R1 units (also applicable to Go2 and B2 legged robots), any attacker within Bluetooth range achieves root code execution$^{\ref{fn:unipwn}}$.
    \item \textbf{Persistent telemetry violates data sovereignty.} MQTT connections to servers at 43.175.228.18:17883 and 43.175.229.18:17883 transmit sensor fusion data at 1.03~Mbps and 0.39~Mbps respectively, with auto-reconnect ensuring continuous surveillance.
    \item \textbf{Humanoid robot platform represents a bidirectional attack vector.} The G1's compromised cryptography and network exposure enable both remote exploitation for surveillance/control and deployment as a mobile cyber-physical weapon platform capable of lateral movement within air-gapped facilities.
    \item \textbf{Cybersecurity AI demonstrates autonomous exploitation capability.} The CAI framework successfully identified and prepared exploitation of authentication bypass vulnerabilities, showcasing the platform's potential as an offensive cyber weapon.
\end{itemize}

The remainder of the brief walks through the platform's anatomy, BLE provisioning vulnerabilities, the FMX cryptanalytic analysis, the telemetry pipeline, and validated attack vectors before closing with what the dual-threat reality means.


\section{Platform Anatomy}

\subsection{Hardware Baseline}
Hardware analysis reveals the G1's compute platform centers on a Rockchip RK3588 SoC (quad-core Cortex-A76 @ 2.4GHz + quad-core Cortex-A55 @ 1.8GHz) with 8GB LPDDR4X RAM and 32GB eMMC storage. Physical security weaknesses include exposed JST debug connectors, unpopulated JTAG pads, and accessible UART interfaces operating at 115200 baud. The sensor array comprises Intel RealSense D435i depth cameras (1280×720 @ 30fps depth, 1920×1080 @ 30fps RGB), dual MEMS microphones, 9-axis IMU, and GNSS receiver—all publishing to DDS topics on the local network. Some of these sensors are susceptible to established attack vectors including optical channel exploitation~\cite{petit2015remote} and acoustic manipulation~\cite{son2015rocking}.

\subsection{BLE Provisioning Vulnerability}
The robot's Bluetooth Low Energy provisioning interface, accessible on models G1, H1, and R1 (also applicable to Go2 and B2 legged robots), contains a critical command injection vulnerability. The Wi-Fi configuration protocol accepts SSID and password fields without sanitization, allowing attackers to inject shell commands using the pattern \textcolor{unitree_primary}{\texttt{";\$(command);\#"}}. All BLE traffic uses AES-CFB encryption with a static key (\textcolor{unitree_primary}{\texttt{df98b715d5c6ed2b25817b6f2554124a}}) and IV (\textcolor{unitree_primary}{\texttt{2841ae97419c2973296a0d4bdfe19a4f}}) hardcoded across the entire fleet, reducing effective entropy to zero. Exploitation requires only BLE proximity and knowledge of these universal credentials, enabling remote code execution with root privileges through the provisioning daemon. The vulnerability persists across firmware versions, with successful exploitation granting persistent access via SSH, credential modification, or arbitrary command execution$^{\ref{fn:unipwn}}$.

\subsection{Service Orchestration}
The 9.2~MB \texttt{master\_service} binary supervises 26 daemons grouped into priority, initialisation, and runtime pools. Runtime profiling recorded \texttt{ai\_sport} consuming roughly 145\% CPU (multi-core), \texttt{state\_estimator} near 30\%, with persistent load from \texttt{robot\_state}, \texttt{vui\_service}, \texttt{webrtc\_bridge}, and \texttt{chat\_go}. Their configuration archives are wrapped by FMX and therefore inherit the weaknesses documented in Section~\ref{sec:fmx}.

\subsection{Communication Surfaces}
Multiple live channels operate simultaneously:
\begin{itemize}
    \item \textbf{DDS/RTPS} for intra-robot publish/subscribe messaging (unencrypted and susceptible to local interception).
    \item \textbf{MQTT} on port~17883 for telemetry and OTA coordination with 43.175.228.18 and 43.175.229.18.
    \item \textbf{WebRTC} via \texttt{webrtc\_bridge} for media streaming, with TLS certificate verification disabled in the shipped client.
    \item \textbf{BLE/Wi-Fi} links via \texttt{upper\_bluetooth} and \texttt{chat\_go} for mobile or app-based control. The BLE channel specifically exposes the command injection vulnerability.
\end{itemize}

The overlap between unencrypted local buses and authenticated cloud uplinks enables the cross-layer attacks explored later in this brief.

\section{FMX Cryptanalysis}\label{sec:fmx}

\begin{figure}[h!]
  \centering
  \includegraphics[width=\columnwidth]{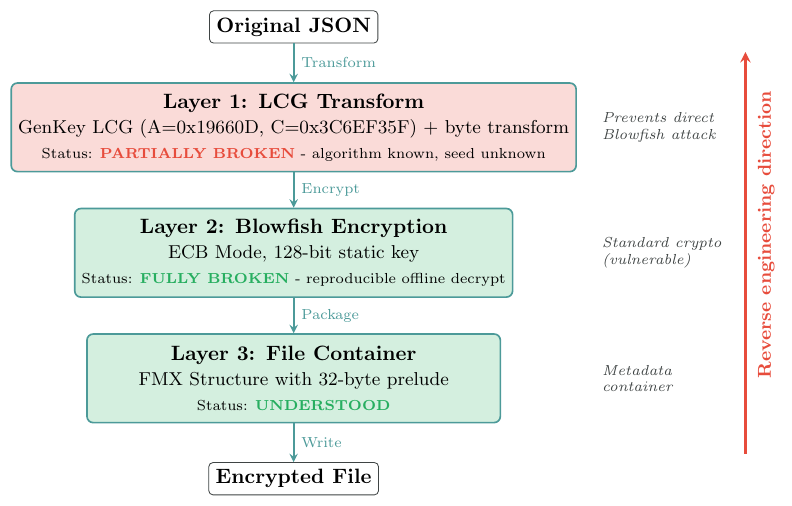}
  \caption{FMX encryption layers reverse-engineered showing partial compromise of Layer 1 (LCG transform) and full compromise of Layer 2 (Blowfish-ECB).}
  \label{fig:encryption}
\end{figure}

Designated “FMX”, the Unitree's encryption employs a dual-layer architecture for configuration protection. Our partial reverse engineering revealed the outer layer uses Blowfish-ECB with a fleet-wide static key, while the inner layer implements a Linear Congruential Generator (LCG) for obfuscation as depicted in Figure \ref{fig:encryption}. While we successfully broke the outer encryption layer and identified the LCG algorithm parameters, complete reversal of the seed derivation mechanism remains unfinished.

\subsection{Layer 2 (Outer): Static Blowfish Encryption—FULLY BROKEN}
\begin{itemize}
\item \textbf{Algorithm:} Blowfish cipher in ECB mode (64-bit blocks, no IV).
\item \textbf{Key Recovery:} Static 128-bit key extracted from \texttt{master\_service} binary through symbol analysis of \texttt{unitree::security::Mixer} class.
\item \textbf{Security Impact:} Fleet-wide key reuse confirmed across multiple G1 units reduces effective entropy to 0 bits—compromising one device exposes the entire fleet. ECB mode enables pattern analysis attacks and provides no authentication guarantees.
\end{itemize}

\begin{center}
\fcolorbox{unitree_primary!30}{cai_light!70}{%
\begin{minipage}{0.9\columnwidth}
\scriptsize\ttfamily\color{unitree_primary}
\noindent Key: 44c56a97ccf33d585a91c18e1c72382b\\
Mode: ECB (no IV), Status: COMPROMISED
\end{minipage}%
}
\end{center}

\subsection{Layer 1 (Inner): LCG Obfuscation—PARTIALLY BROKEN}
\begin{itemize}
\item \textbf{Algorithm:} Linear Congruential Generator with parameters matching glibc's \texttt{rand()}: $X_{n+1} = (1664525 \cdot X_n + 1013904223) \bmod 2^{32}$ (where 1664525 = 0x19660D, 1013904223 = 0x3C6EF35F).
\item \textbf{Key Extraction:} High byte of each LCG state used as XOR key: \texttt{key[i] = (state >> 24) \& 0xFF}.
\item \textbf{Seed Derivation:} While we identified the LCG algorithm, the exact seed derivation from device identifiers remains partially understood. The 32-bit seed space is tractable for brute force but was not fully explored.
\item \textbf{Current Status:} Algorithm reverse-engineered, parameters identified, seed generation incompletely documented.
\end{itemize}




\section{Telemetry Exposure}

\subsection{Captured Data Surface}
Our SSL\_write instrumentation on 9~September~2025 captured a 10-minute telemetry session, revealing structured JSON payloads of 4.5--4.6~KB transmitted at 300-second intervals. Each frame contained:
\begin{itemize}
    \item \textbf{Battery telemetry} (cell voltages, currents, temperatures, SoC).
    \item \textbf{IMU orientation} (pitch, roll, yaw) alongside torque and temperature from every joint.
    \item \textbf{Service inventories} reporting enablement state for motion, voice, perception, and OTA modules.
    \item \textbf{Resource metrics} (CPU load arrays, memory usage, filesystem statistics).
\end{itemize}

The telemetry architecture employs a dual-channel design: periodic MQTT state reports (300-second intervals) complement continuous DDS streams carrying real-time sensor data. The DDS topics including audio (\texttt{rt/audio\_msg}), video (\texttt{rt/frontvideostream}), LIDAR point clouds (\texttt{utlidar/cloud}), and proprioceptive feedback enable passive extraction of this data by simply listening to network traffic on the local network segment.

\subsection{Exfiltration Channels and Throughput}
Two MQTT endpoints anchor the exfiltration pipeline:
\begin{enumerate}
    \item \textbf{43.175.228.18:17883}: primary broker for \texttt{robot\_state\_service} (approximately 1.03~Mbps during capture).
    \item \textbf{43.175.229.18:17883}: secondary broker for \texttt{ota\_boxed} failover (approximately 0.39~Mbps steady state).
\end{enumerate}

The \texttt{chat\_go} service maintains a WebSocket connection to 8.222.78.102:6080 with SSL verification disabled, enabling conversational transcripts and control messages to transit outside the MQTT path. All connections auto-recover within seconds of disruption, reinforcing the platform’s always-on data posture.

\subsection{Legal and Sovereignty Exposure}
Streaming multi-modal telemetry to Chinese infrastructure invokes that country’s cybersecurity law, mandating potential state access. In European deployments the absence of consent or transparency mechanisms contravenes GDPR Articles~6 and~13, while California operators lack the opt-out required by the CCPA. Sensitive facilities face an additional national-security risk: continuous audio, video, spatial maps, and actuator states can be relayed off-site in real time with no user awareness. 


\section{Attack Vectors: Dual-Threat Humanoid Platforms}

The G1 ultimately behaved as a dual-threat platform: covert surveillance at rest, weaponised cyber operations when paired with the right tooling. Two case studies illustrate the spectrum of risk.

\subsection{Case Study 1: Humanoids as Surveillance Trojan Horses}

Network analysis conducted in September~2025 documented the G1's persistent connectivity behavior: MQTT client processes (\texttt{robot\_state\_service} PID 1226, \texttt{ota\_boxed} PID 691) establish TLS 1.3 connections to servers at 43.175.228.18:17883 and 43.175.229.18:17883 within 5 seconds of system initialization. Our SSL\_write instrumentation on 9~September captured telemetry transmissions at 300-second intervals (configured via \texttt{ReportInterval} parameter), each containing comprehensive system state.

While our limited telemetry inspection didn't allow us to confirm this, we observed the following surveillance channels in the robot's computational graph:

\textbf{Observed surveillance channels:}
\begin{itemize}
\item \textbf{Audio:} Continuous capture via \texttt{vui\_service} through dual microphones, streaming to \texttt{rt/audio\_msg} DDS topic without user indicators
\item \textbf{Visual:} RealSense camera at 1920×1080@15fps with H.264 encoding, cloud streaming via Amazon Kinesis SDK
\item \textbf{Spatial:} LIDAR point clouds (\texttt{utlidar/cloud}), 3D voxel mapping, GPS/GNSS positioning with sub-centimeter odometry tracking
\end{itemize}

Given the covert nature of the robot data collection, we argue that the channels described above could be used to conduct surveillance on the robot's surroundings, including audio, visual, and spatial data. This combination enables silent meeting capture, document imaging, facility mapping, and behavioural profiling—everything needed for corporate espionage—while routing the results offshore without operator awareness.

\subsection{Case Study 2: Weaponized Cybersecurity AI Platform}

We deployed the Alias Robotics Cybersecurity AI (CAI) framework~\cite{aliasrobotics2025cai,cai2025github} directly on the G1's Rockchip RK3588 processor to evaluate autonomous exploitation capabilities. The CAI agent, building upon LLM-powered penetration testing approaches~\cite{deng2023pentestgpt}, executed a systematic four-phase assessment—reconnaissance, vulnerability analysis, exploitation preparation, and attack surface mapping—producing the following findings:
\begin{itemize}
\item \textbf{Recon.} Enumerated live connections and identified MQTT/WebSocket/WebRTC endpoints reachable from inside the robot. Additionally confirmed the BLE provisioning vulnerability allows trivial initial compromise using hardcoded AES keys.
\item \textbf{Exploitation rehearsal.} Attempted broker logins using the recovered credentials (anonymous access rejected, certificate-authenticated sessions prepped but withheld during ethical testing), demonstrating the feasibility of command injection via both BLE provisioning and direct telemetry spoofing.
\item \textbf{Attack planning.} Produced an attack surface matrix covering BLE-based command injection, MQTT topic abuse, WebRTC stream hijack, and OTA manipulation pathways.
\end{itemize}

The CAI leveraged the robot's authenticated position within Unitree's infrastructure to enumerate attack vectors without triggering defensive mechanisms. This demonstration validates the dual-use risk: platforms designed for legitimate telemetry can be repurposed for offensive operations. The autonomous nature of CAI-driven attacks—operating at machine speed without human intervention—necessitates equivalent defensive capabilities.




\section{Conclusion: The Dual Threat and the Cybersecurity AI Imperative}

\noindent This paper presents a systematic security assessment of the Unitree G1 humanoid robot platform, demonstrating its dual functionality as both a surveillance vector and an active cyber operations platform. Through static analysis and runtime behavioral observation, we conducted partial reverse engineering of Unitree's proprietary FMX encryption scheme, enabling comprehensive evaluation of the platform's security architecture. Our analysis indicates this represents the most sophisticated security implementation observed in commercial robotics platforms to date, much more mature than the industry average as depicted in Table \ref{tab:comparison}.

\begin{table}[H]
  \centering
  \footnotesize
  \begin{tabular}{lcc}
  \toprule
  \textbf{Feature} & \textbf{G1 Robot} & \textbf{Industry Average} \\
  \midrule
  Encrypted Config & \textcolor{unitree_green}{\checkmark} & \textcolor{unitree_red}{\texttimes} \\
  Dynamic Credentials & \textcolor{unitree_green}{\checkmark} & \textcolor{unitree_red}{\texttimes} \\
  Hardware Binding & \textcolor{unitree_green}{\checkmark} & \textcolor{unitree_orange}{$\sim$} \\
  Multi-layer Defense & \textcolor{unitree_green}{\checkmark} & \textcolor{unitree_red}{\texttimes} \\
  No Hardcoded Secrets & \textcolor{unitree_green}{\checkmark} & \textcolor{unitree_red}{\texttimes} \\
  \bottomrule
  \end{tabular}
  \caption{\footnotesize Security feature comparison with industry standards.}
  \label{tab:comparison}
\end{table}

Despite this robust security posture, we identified vulnerabilities that enable the robot to function as an attack vector, validated through two empirical case studies: (a) trojan horse deployment for covert data exfiltration and (b) platform compromise for lateral movement operations. Network traffic analysis via SSL\_write capture over a 10-minute observation period confirmed continuous telemetry transmission to remote servers (43.175.228.18:17883 and 43.175.229.18:17883), transmitting multi-modal sensor data and service state information at 300-second intervals without explicit user consent or notification mechanisms. This data collection architecture presents potential violations of privacy regulations including GDPR Articles~6 (lawfulness of processing) and 13 (information to be provided). Beyond passive surveillance documentation, we operationalized a Cybersecurity AI (CAI) agent on the Unitree G1 platform to conduct reconnaissance and prepare exploitation vectors against the manufacturer's cloud infrastructure, demonstrating escalation pathways from covert data collection to active offensive cyber operations. Our findings indicate that securing humanoid robots requires fundamental paradigm shifts toward adaptive Cybersecurity AI frameworks~\cite{cai2025github} capable of addressing the unique challenges inherent in physical-cyber convergence systems. This research contributes empirical evidence critical for developing comprehensive security standards~\cite{surve2024sok,sae2021j3016} as humanoid robots transition from research platforms to operational deployment in critical infrastructure domains.

\section*{Acknowledgments}

CAI was developed by Alias Robotics and co-funded by the European Innovation Council (EIC) as part of the accelerator project "RIS" (GA 101161136) - HORIZON-EIC-2023-ACCELERATOR-01 call.

\vspace{0.3cm}
\bibliography{bibliography}

\vspace{0.5cm}
\noindent\rule{\textwidth}{0.4pt}
\vspace{0.3cm}

\small
\noindent\textbf{Disclosure:} Findings reported to manufacturer prior to publication.\\
\end{document}